# Advanced RF Structures for Wakefield Acceleration and High-Gradient Research

Xueying Lu,[i,ii] Jiahang Shao,[i,xv] John Power,[i] Chunguang Jing,[i,iii] Gwanghui Ha,[i] Philippe Piot,[i,ii] Alexander Zholents,[i] Richard Temkin,[iv] Michael Shapiro,[iv] Julian Picard,[iv, xvi] Bagrat Grigoryan,[v] Chuanxiang Tang,[vi] Yingchao Du,[vi] Jiaru Shi,[vi] Hao Zha,[vi] Dao Xiang,[vii] Emilio Nanni,[viii] Brendan O'Shea, [viii] Yuri Saveliev,[ix] Thomas Pacey,[ix] James Rosenzweig,[x] Gerard Andonian,[x] Evgenya Simakov,[xi] Francois Lemery, [xii] Alex Murokh,[xiii] Sergey Kutsaev,[xiii] Alexander Smirnov,[xiii] Ronald Agustsson,[xiii] Gongxiaohui Chen,[i] Seunghwan Shin,[xiv] Ben Freemire [iii]

i      Argonne National Laboratory, Lemont, IL, USA
ii     Northern Illinois University, DeKalb, IL, USA
iii    Euclid Techlabs LLC, Bolingbrook, IL, USA
iv     Massachusetts Institute of Technology, Cambridge, MA, USA
v      CANDLE SRI, Yerevan, Armenia
vi     Tsinghua University, Beijing, China
vii    Shanghai Jiao Tong University, Shanghai, China
viii   SLAC National Accelerator Laboratory, Menlo Park, CA, USA
ix     Cockcroft Institute, Sci-Tech Daresbury, Warrington, UK
x      University of California, Los Angeles, CA, USA
xi     Los Alamos National Laboratory, Los Alamos, NM, USA
xii    Deutsches Elektronen-Synchrotron DESY, Hamburg, Germany
xiii   RadiaBeam Technologies, Santa Monica, CA, USA
xiv    Pohang University of Science and Technology, Pohang, Gyeongbuk, South Korea
xv     Currently at Institute of Advanced Science Facilities, Shenzhen, China
xvi    Currently at SpaceX

## Table of Contents

## Executive summary

Structure wakefield acceleration (SWFA) is one of the most promising AAC schemes in several recent strategic reports, including DOE's 2016 AAC Roadmap, report on the Advanced and Novel Accelerators for High Energy Physics Roadmap (ANAR), and report on Accelerator and Beam Physics Research Goals and Opportunities. SWFA aims to raise the gradient beyond the limits of conventional radiofrequency (RF) accelerator technology, and thus the RF to beam energy efficiency, by reducing RF breakdowns from confining the microwave energy in a short (on the order of about 10 ns) and intense pulse excited by a drive beam.

We envision that the following research topics, within the scope of AF7, are of great interest in the next decade:

### (1) Advanced wakefield structures

SWFA requires optimization of unique structures to achieve high gradients and high efficiencies. Advanced structures with novel electromagnetic characteristics can dramatically improve SWFA performance by pushing the boundaries in the structure design. Much progress has been made in recent years. Dielectric structures (including dielectric tubes, dielectric slabs and dielectric disk-loaded structures) and metallic periodic structures in both X-band and in the millimeter wave band. Structures with novel topologies, such as metamaterial (MTM) structures, photonic bandgap (PBG) structures and photonic topological crystals, have been successfully been tested.

### (2) Terahertz and sub-terahertz (THz) structures

SWFA in the THz regime, using short RF pulses, is a promising research direction to pursue. THz structures have the advantages of strong beam-structure interaction (high shunt impedance and thus high gradient) and small transverse size, which could lead to compact and cost-effective future colliders. When combined with bunch shaping techniques, THz structures could be ideal for high-gradient and high-efficiency wakefield acceleration. Recent advances in fabrication have made it possible to push SWFA into the mm-wave and THz regime.

### (3) RF breakdown physics

The physics of RF breakdown has been highly sought-after in the high-gradient acceleration community. Previous RF breakdown studies were mostly carried out on conventional RF linac structures, with pulse lengths ranging from a few hundred nanoseconds to a few microseconds. Much of our current knowledge about breakdowns in the short-pulse high-frequency regime is through extrapolation of data obtained at other conditions. Early SWFA experimental evidence has shown that short-pulse operation (a few ns) has the potential to dramatically increase the gradient. Further research on this topic could bring valuable insight to RF breakdown physics.

**Long-term applications and synergies with other AAC concepts**

Research on SWFA in the above directions would directly contribute to long-term large-scale applications, including AAC-based linear colliders and compact light sources. Integrated studies with beam shaping and control techniques are needed towards the realization of these large-scale applications. There is also potentially a strong synergy between SWFA and other AAC concepts, when structures are combined with plasmas into hybrid AAC schemes.

To conclude, research on novel structures is at the core of advancing SWFA, and is critical to future AAC-based linear colliders; at the same, it has a strong synergy with other directions, such as cavity designs, high-power microwave systems and sources, and compact light sources.

# 1. Introduction

The advanced accelerator concept (AAC) community conducts long-term research aimed at future energy-frontier colliders. AAC-based future colliders have the potential to operate at substantially higher energy and lower cost beyond the capability of current accelerator technologies. Among these advanced concepts, structure wakefield acceleration (SWFA) is one promising candidate for a TeV scale compact linear collider. In SWFA, a high-charge drive beam traverses a structure in vacuum and excites intense wakefield. The generated wakefield can be used to accelerate a low-charge main beam (or witness beam), in either the same structure (collinear wakefield acceleration, or CWA) or a separate structure in parallel (two-beam acceleration, or TBA). The two schemes of SWFA, with many similarities, have complementary strengths for different applications.

Compared to conventional radiofrequency (RF) accelerators powered by klystrons, SWFA employs short electron bunches as compact power sources. It is promising to raise the gradient, which is limited to about 100 MV/m in conventional RF accelerators due to RF breakdowns, by confining an intense wakefield in short RF pulses. Based on the extensive research in the high-gradient acceleration community, using short RF pulses could mitigate the risk of RF breakdown and increase the operating gradient as a result [2].

Linear colliders and compact light sources based on SWFA have been studied and proposed. The Compact Linear Collider [3,4] is the most mature SWFA-based linear collider design, as a TeV-scale collider based on the TBA scheme, with a loaded gradient of about 100 MV/m, and an RF pulse length of 240 ns. The Argonne Flexible Collider, proposed is proposed to operate with a shorter RF pulse of 20 ns, more than an order of magnitude shorter than those typically used in pulsed high-power microwave sources, to achieve an ~300 MV/m gradient [5, 6]. Under the CWA concept, a compact multi-user X-ray free electron laser (XFEL) has been proposed [7], where CWA accelerating modules are utilized for high-gradient high-efficiency acceleration before the undulator sections.

Research and development on advanced structures are critical to the success of the proposed SWFA-based large-scale machines. This White Paper will focus on the recent progress and future directions regarding advanced RF structures for SWFA, and is organized as follows. In Section 2, we summarize the recent progress in the SWFA community on advanced RF structures and high-gradient research, and we also identify future directions in the next decade, which could have an impact in the AAC community and the broader high energy physics (HEP) community. In Section 3, we discuss the long-term large-scale applications, including future AAC-based linear colliders and compact light sources. We also discuss potential interests in developing hybrid AAC schemes combining SWFA and plasma technologies. We present concluding remarks in Section 4.

## 2. Recent Advances and Future Directions

We summarize recent advances and future directions on the following topics: (2.1) advanced wakefield structures; (2.2) Terahertz and Sub-terahertz (THz) Structures; (2.3) RF breakdown physics with short pulses.

### 2.1 Advanced Wakefield Structures

Research on advanced structures is at the core of SWFA research. SWFA requires different structure optimization from the unique gradient and efficiency considerations associated with short RF pulses. Desired features for short-pulse SWFA include low RF loss, high group velocity, and high shunt impedance. In conventional metallic disk-loaded structures, there is a tradeoff that a high group velocity would lead to a low shunt impedance. Therefore, advanced wakefield structures tailored specifically for the SWFA application are needed. Several promising advanced wakefield structures have been studied or are under investigation. Figure 1 presents a few examples.

The dielectric-loaded accelerators (DLAs) have been studied due to its low cost and ease of fabrication [8],

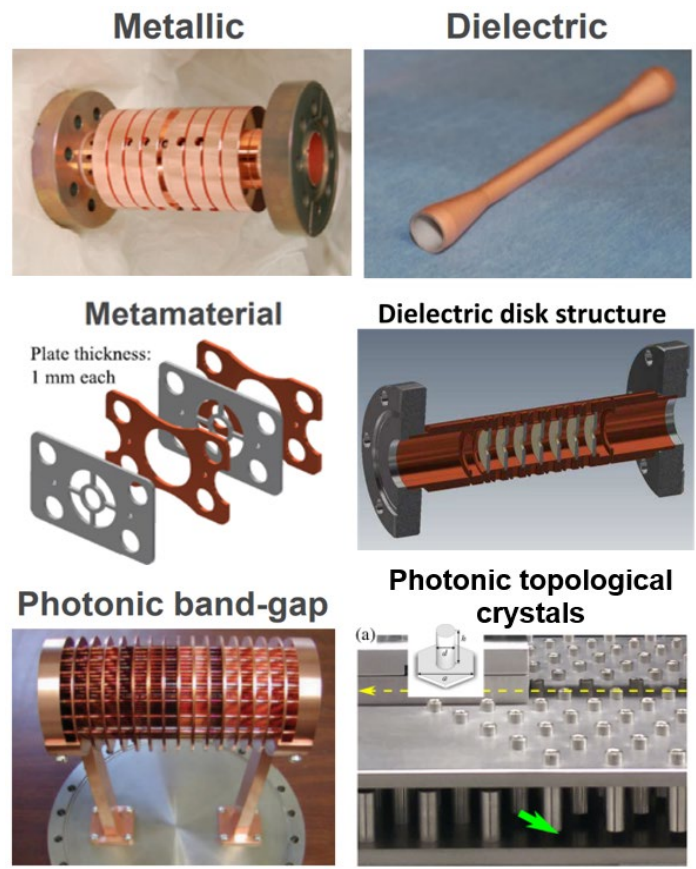

*Figure 1: Examples of advanced structures studied and tested for wakefield acceleration.*

and 200 MW extracted power and sub-GV/m gradient have been achieved [9]. In DLAs, the structure used to interact with the relativistic beam is often a dielectric tube with copper coating. The RF loss in DLAs can be potentially reduced by using ultralow-loss dielectric materials [10].

Metallic disk-loaded structures [11-13] have be demonstrated as wakefield power extractors, with up to 400 MW peak power generated [12], and as accelerators in the TBA regime, with a gradient of up to 300 MV/m achieved [13].

In recent years, structures with novel topologies have been explored.

Metamaterial (MTM) structures [14-16], as subwavelength periodic structures with novel electromagnetic properties not found in nature, have been demonstrated as successful power extractors. Compared to conventional structures, the negative group velocity in MTM structures enables simultaneous high shunt impedance and high group velocity. In a series of experiments since 2018, a record-high power of 565 MW [16] has been achieved in an X-band MTM structure. Ongoing research is on applying MTM structures to complete high-gradient TBA.

Dielectric disk-loaded accelerators (DDAs) [17, 18] have also been proposed and tested as promising candidates for short-pulse TBA. DDA structures employ low-loss dielectric disks between cylindrical cells to enable higher shunt impedance and higher group velocity than DLA structures. In comparison, DDA structures are easier to fabricate and to tune, and can provide a higher efficiency. Several DDA structures are under testing at AWA as accelerator structures for TBA.

Other advanced structures investigated in the community include the photonic bandgap structures [19] to suppress higher order modes while confining only the fundamental mode, topological crystals [20], woodpile structures [21], and bimodal cavities [22]. Parallel-coupled structures with a short filling time are being developed for ultrashort input power pulses [23], and could be applied to SWFA to achieve high gradients with pulses of a few nanoseconds long.

Progress on advanced wakefield structures could improve the RF-to-beam efficiency by 50% and therefore reduce the site power and operation costs.

**Future directions in advanced wakefield structures include:**

(1) Identifying new concepts with novel structure topologies and coupling schemes in support of short-pulse SWFA; (2) Improving the power and gradient achieved in structure candidates, such as DLA, MTM, and DDA through RF design and beamline optimization; (3) Exploring fabrication techniques to obtain higher reliability and reduce the cost of structure fabrication; (4) Investigating higher-order mode damping methods to mitigate long-range transverse wakefield; (5) Building fully-functional demonstrators for TeV-scale colliders.

## 2.2 Terahertz and Sub-terahertz (THz) Structures

Terahertz and sub-terahertz (THz) structures have seen much progress in recent years [7, 24-32]. THz structures are known to have a strong beam-structure interaction and a small transverse size, so that a high gradient could be achieved in compact accelerating structures. These advantages also apply to SWFA in the THz frequency. THz structures for SWFA have become more desirable due to rapid advances in advanced fabrication techniques [28-33], such as additive manufacturing, electroforming, two-half assembly without brazing, and precise etching control of dielectric.

Figure 2 presents a few examples of recent demonstrations of beam-driven THz acceleration, with the frequency ranging from W-band to about 400 GHz.

Dielectric lined-waveguides at about 400 GHz [27] have been excited with electron bunches at FACET, with an energy of about 20 GeV. High acceleration gradient of above 300 GeV/m was measured.

Metallic two-half structures have been studied at both FACET [28, 29] and AWA [30]. The FACET two-half structure, when stimulated by the 20 GeV beam, achieved a maximum gradient of 0.3 GV/m, with a peak surface electric field of 1.5 GV/m and a RF pulse length of about 2.4 ns [29]. The AWA structure was excited by a train of three 2 nC bunches, separated at 1.3 GHz, with an energy of 65 MeV [30]. A 5 MW rf pulse at 91 GHz was generated, and the corresponding accelerating gradient was 85 MV/m.

Metallic corrugated waveguide structures at about 180 GHz have been proposed to serve as compact CWA-based accelerators for multi-beamline XFEL [7]. Such waveguides have been built at ANL and tested at Brookhaven National Laboratory's Accelerator Test Facility (ATF) [31], where the wakefield was characterized in the THz corrugated waveguide when it was excited by the ATF 55 MeV electron beam with a bunch charge of about 150 pC.

Recently at AWA, metallic structures at about 200 GHz fabricated by the Pohang University of Science and Technology (POSTECH) have been utilized for longitudinal and transverse wakefield mapping. In a theoretical study [34], 1.4 THz structures are proposed to provide GV/m gradient when driven by THz bunch trains.

THz structures have great potentials to enable high-gradient high-efficiency wakefield acceleration, when there is a combined effort of structure R&D and precise phase-space control of

Dielectric-lined waveguide (~400 GHz)

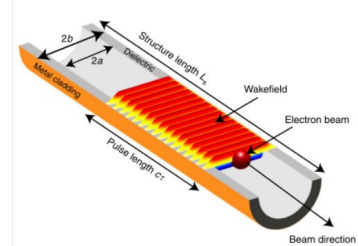

Metallic two-half structure (~120 GHz to 140 GHz)

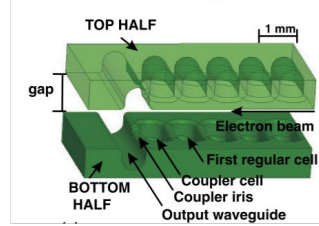

Metallic two-half structure (91 GHz)

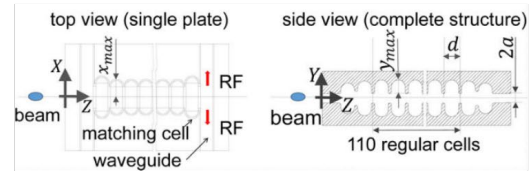

Metallic corrugated waveguide (~180 GHz)

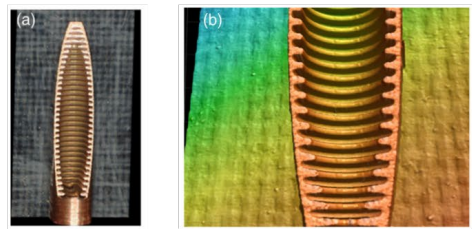

*Figure 2: Examples of THz structures for SWFA. From top to bottom, the four figures are from Ref. [27], [28], [30] and [31], respectively.*

the electron beam. Asymmetric current distributions can enhance the transformer ratio (defined as the ratio of the accelerating gradient at the witness/main beam location and the decelerating gradient at the drive beam location) beyond the theoretical limit of 2 for symmetric bunches [35]. Proper beam shaping can also extend the interaction length of the drive bunch in the structure and increase the total extractable energy as a result.

**Future directions in THz wakefield structures include:**

(1) Fabricating matched power extractor (decelerator) and accelerator to demonstrate the TBA concept at high frequencies; (2) Studying advanced approach to improve the power coupling between decelerator and accelerator; (3) Developing collider concept based on high-frequency short-pulse TBA; (4) Studying short-range and long-range BBU control in THz structures; (5) Combining THz wakefield structure research and development with bunch shaping techniques to enable high-gradient high-efficiency wakefield acceleration.

### 2.3 RF Breakdown Physics with Short Pulses

RF breakdown poses a fundamental limit on achievable gradient in accelerator structures. Previous research reveals that the breakdown rate is lower with a short RF pulse duration [2], which implies that short-pulse SWFA with a RF pulse length of a few nanoseconds could potentially achieve higher gradient.

Research on RF breakdown in short-pulse SWFA is complementary to previous and ongoing breakdown studies in the high-gradient community, where extensive research has been conducted to understand the various breakdown mechanisms. Different approaches have been investigated to mitigate the RF breakdown risk, such as cryogenic operation [36], curved surface to reduce internal multipacting current [37], and multipacting suppression by external magnetic field [38].

At the same time, SWFA could open up a new parameter space for breakdown characterization, with short RF pulses on the order of a few nanoseconds and high frequencies. Most of previous research in the community focuses on long-pulse operation (>100 ns); as a result, information on short-pulse operation (~10 ns) is based on theoretical assumptions and extrapolations. Therefore, there is an urgent need to investigate the structure limitation in the short-pulse regime with direct high-power tests. Such tests could in return bring new insights to the fundamental physics of RF breakdown.

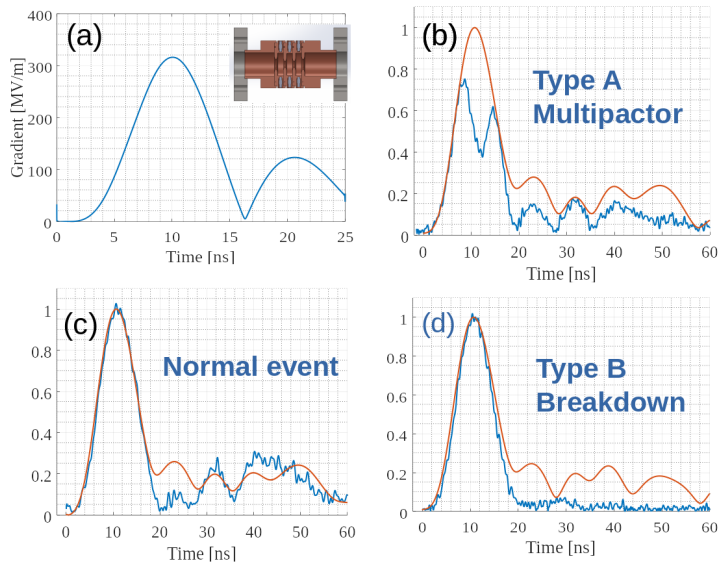

*Figure 3 Experimental tests [13] performed on a 3-cell accelerating structure with a short RF input pulse showing the predicted time-dependent accelerating gradient, as in (a), with measured transmitted power, as in (b) Type A multipactor, (c) normal operation, and (d) Type B breakdown. The orange trace corresponds to numerical calculations performed with CST microwave studio while the blue traces are measured waveforms.*

In a recent experiment at AWA [13] on an X-band power extraction and transfer structure (PETS) for SWFA, indirect evidence was collected on a potential new regime of acceleration, as the Breakdown Insensitive Acceleration Regime (BIAR). In the BIAR regime, a breakdown event does not disrupt the accelerating field in the primary pulse due to the short pulse length. Figure 3 presents some RF pulses collected during the experiment. As can be seen in Fig. 3 (d), the "Type B" breakdown in the BIAR regime does not alter the primary pulse for acceleration. Further investigation is needed to understand this regime.

**Future directions in RF breakdown studies for SWFA include:**

(1) Studying the gradient scaling law in a short-pulse high-frequency regime, which is complementary to breakdown studies in klystron-powered accelerating structures; (2) Demonstrating high-gradient structures which take full advantage of recent discoveries of the limiting factors; (3) Continuing fundamental research to understand the physics behind the limiting factors, and proposing reliable methods to overcome them, such as the Breakdown Insensitive Acceleration Regime (BIAR).

## 3. Linear Collider, Light Sources, and Other Synergies

### 3.1 Argonne Flexible Linear Collider (AFLC)

Details about the proposed 3 TeV Argonne Flexible Linear Collider (AFLC) can be found in another White Paper, entitled "Continuous and Coordinated Efforts of Structure Wakefield Acceleration (SWFA) Development for an Energy Frontier Machine", by Chunguang Jing *et al.*, submitted to AF6 in Snowmass 2021. A 500 MeV TBA demonstrator is also presented in the above-mentioned AF6 White Paper as a near-term deliverable.

### 3.2 Compact Light Sources Based on SWFA

Details about the proposed compact multi-user XFEL based on the CWA concept can be found in Ref. [7] and in another White Paper, entitled "Continuous and Coordinated Efforts of Structure Wakefield Acceleration (SWFA) Development for an Energy Frontier Machine", by Chunguang Jing *et al.*, submitted to AF6 in Snowmass 2021. A THz CWA energy doubler for XFEL is also presented in the above-mentioned AF6 White Paper as a near-term deliverable.

### 3.3 Integrated Studies with Beam Shaping and Dynamics

To realize the full potential of SWFA, integrated studies are needed to combine advanced structure R&D with beam shaping and control techniques. Significant progress has been made in precise control of the electron-beam distribution: several shaping methods have been developed and experimentally demonstrated [39-43]. An analytical theory of the beam breakup (BBU) instability has been formulated [44] and mitigation techniques developed [45], and numerical-simulations validation is underway (i.e., tapered external-focusing configurations were devised) [46]. An important step toward the realization of an SWFA-based large-scale facility is the development of an integrated experiment where shaped drive bunches supporting enhanced transformer ratios are propagated in a meter-scale SWFA module [7] to explore the onset of the BBU instabilities while also supporting the development of associated beam diagnostics and correction techniques.

### 3.4 Hybrid AAC Applications

Hybrid applications have been studied involving SWFA and other AAC concepts; here we list a few examples. The Capillary Trojan Horse (CTH) method [47] has been investigated, where the plasma wakefield accelerator in the Trojan Horse method is replaced with a dielectric accelerator, while the plasma photocathode technique is maintained. A proof-of-concept experiment on the hybrid concept has been performed at AWA in an effort to generate low-emittance beams. There is also a proposal on wakefield excitation by a ramped bunch train in a collinear, single-channel THz DLA structure filled with a low-temperature plasma [48]. This hybrid concept is predicted to improve the transformer ratio while stabilizing the electron bunches transversely in the plasma. Similar hybrid AAC applications would be interesting to pursue in the future.

## 4. Conclusions

SWFA is a promising candidate for future AAC-based linear colliders and other applications. Compared to conventional klystron-powered acceleration techniques, short-pulse SWFA has great potentials to achieve a dramatically higher accelerating gradient, which could lead to compact TeV scale colliders. Since SWFA requires specialized structure optimization, research on advanced wakefield structures is critical, and has seen much progress in recent years. Pushing SWFA into the THz frequency range also has clear advantages. When combined with precise bunch shaping techniques, THz SWFA could enable high-gradient high-efficiency acceleration. Exploring the gradient limit of short-pulse SWFA could complement our existing knowledge on RF breakdowns, and potentially reveal new physics in the ultrashort pulse length parameter space. Advanced RF structures for wakefield acceleration has strong synergies with future AAC-based linear colliders (such as CLIC and AFLC), and compact light sources (such as CWA-based XFELs). SWFA research also has strong ties with the RF subgroup in AF7, on topics including cavity designs, high-power microwave systems and sources.